\documentclass{ws-procs9x6}

\begin{document}

\title{BIPARTITE-FINSLER SPACES AND THE BUMBLEBEE MODEL}

\author{J. EUCLIDES G. SILVA$^*$ and C.A.S. ALMEIDA}

\address{Physics Department, Cear\'{a} Federal University ,\\
Fortaleza, Cear\'{a} ZIP 6030, 60455-760, Brazil\\
$^*$E-mail: euclides@fisica.ufc.br\\
}



\begin{abstract}
We present a proposal to include Lorentz-violating effects in
gravitational field by means of the Finsler geometry. In the Finsler set up, the
length of an event depends both on the point and the direction in the space-time. We
briefly review the bumblebee model, where the Lorentz violation is induced by
a spontaneous symmetry breaking due to the bumblebee vector field.The main geometrical
concepts of the Finsler geometry are outlined. Using a Finslerian Einstein-Hilbert action
we derive the bumblebee action from the bipartite Finsler function
with a correction to the gravitational constant.
\end{abstract}

\bodymatter

\section{Bumblebee model}

A model to include gravity in the Standart model extension (SME) \cite{Colladay:1996iz} is provided by a vector field
$B_{\mu}$, the so-called bumblebee field, which couples with the usual geometrical tensors through the action \cite{Kostelecky:2003fs}
\begin{equation}
 \label{bumblebeeactioninteraction}
 S_{LV}=K\int_{M}{(uR + s^{\mu\nu}R_{\mu\nu})\sqrt{-g}d^{4}x}
\end{equation}
where $ s^{\mu\nu} =  \xi\left(B^{\mu}B^{\nu}- \frac{1}{4}B^{2}g^{\mu\nu}\right)$
and $u =  \xi B^{\mu}B_{\mu}.$
Inspite this model introduces the Lorentz violation into the geometry \cite{Maluf:2013nva}, the geometrical
tensors remain Lorentz-invariants. This drawback can be overcame by means of the Finsler geometry \cite{Kostelecky:2003fs}
.

\section{Finsler geometry}

A Finsler geometry is an extension of the Riemannian geometry where
given a curve $\gamma:[0,1]\rightarrow M$, its arc length is given by \cite{chern}

\begin{equation}
 s=\int_{0}^{1}{F(x,y)dt},
\end{equation}
where, $x\in M, y\in T_{x}M$. The function $F(x,y)$ is called the Finsler principal function.
Note that the interval depends both on the position $x$ as on the direction $y$.

As in the Riemannian case,
it is possible to define a Finsler metric by \cite{chern}
\begin{equation}
 \label{finslermetric}
 g^{F}_{\mu\nu}(x,y)=\frac{1}{2}\frac{\partial^{2}F^{2}}{\partial y^{\mu}y^{\nu}}.
\end{equation}

The Finsler metric (\ref{finslermetric}) is a symmetric and an anisotropic quadratic form on $TTM$.
Differenting the metric yields the so-called Cartan tensor $A_{\alpha\beta\gamma}(x,y)=\frac{F}{4}\frac{\partial g_{\alpha\beta}}{\partial y^{\gamma}}
$ from which it is possible to define a nonlinear connection by
$N_{\delta}^{\alpha}= \gamma_{\alpha\beta}^{\delta}y^{\beta} - \frac{A^{\delta}_{\alpha\beta}}{F}\gamma^{\beta}_{\epsilon\xi}y^{\epsilon}y^{\xi}$
which decouples $TTM$ into
$ TTM=hTTM \bigoplus vTTM$, where $\frac{\delta}{\delta x^{\alpha}}=\frac{\partial}{\partial x^{\alpha}} - N^{\beta}_{\alpha}\frac{\partial}{\partial y^{\beta}}
$ is the basis for the horizontal section $hTTM$ and $F\frac{\partial}{\partial y^{\alpha}}$ is a base
for the vertical section $vTTM$.
The compatible Cartan connection is given by $\omega_{\alpha}^{\delta}=\Gamma^{\delta}_{\alpha\beta}dx^{\beta} + \frac{A_{\alpha\beta}^{\delta}}{F}\delta y^{\beta},
$
where
$ \Gamma^{\delta}_{\alpha\beta} = \frac{1}{2}g^{F\delta\epsilon}_{\alpha\beta}(\delta_{\alpha} g_{\epsilon\beta}+\delta_{\beta} g_{\epsilon\alpha}-\delta_{\epsilon} g_{\alpha\beta})
$\cite{chern}.

Following the approach proposed by Pfeifer and Wohlfarth \cite{Pfeifer:2011xi}, here we are concerned with only the Lorentz violation effects on tensor fields defined on $hTTM$.
The horizontal Ricci curvature is defined as \cite{chern}

\begin{equation}
R_{\alpha\beta}  =  \delta_{\gamma}\Gamma^{\gamma}_{\alpha\beta} - \delta_{\beta}\Gamma^{\gamma}_{\alpha\gamma} + \Gamma^{\gamma}_{\epsilon\beta}\Gamma^{\epsilon}_{\alpha\gamma} - \Gamma^{\gamma}_{\epsilon\gamma}\Gamma^{\epsilon}_{\alpha\beta},
\end{equation}
and the scalar curvature $R^{F} = g^{F\alpha\beta}R^{F}_{\alpha\beta}$.


\section{The Bipartite space}

Based in a previous work on the classical point particles Lagrangians \cite{Kostelecky:2010hs}, Kostelecky proposed a new Finsler function of form \cite{Kostelecky:2011qz}

\begin{equation}
\label{kosteleckyfinslerfunction}
 F(x,y)=\sqrt{g_{\mu\nu}(x)y^{\mu}y^{\nu}} + l_{P}(a_{\mu}(x)y^{\mu} \pm\sqrt{s_{\mu\nu}(x)y^{\mu}y^{\nu}}),
\end{equation}
where $s_{\mu\nu}(x)=b^{2}(x)g_{\mu\nu}(x) - b_{\mu}(x)b_{\nu}(x).$
The Planck length scale $l_{P}$ provides a scale of
length where the anisotropic effects have to be taken in account.

Kostelecky, Russel and Tso supposed $a_{\mu}=0$ and enhanced the $s_{\mu\nu}$ tensor to be any symmetrical one \cite{Kostelecky:2012ac}. This geometry is so-called bipartite.

The bipartite-Finsler function yields the Finsler metric \cite{Kostelecky:2012ac}

\begin{equation}
\label{bipartitefinslermetric}
 g_{\mu\nu}^{F}=\frac{F}{\alpha}g_{\mu\nu} + l_{P}^{2}\left(\frac{F}{\sigma}s_{\mu\nu} - \alpha\sigma k_{\mu}k_{\nu}\right),
\end{equation}
where, $k_{\mu}  =  \frac{1}{\alpha}\frac{\partial \alpha}{\partial y^{\mu}} - \frac{1}{\sigma}\frac{\partial \sigma}{\partial y^{\mu}}$.
Furthermore, for $dim M=4$, the relation between the Finslerian and Lorentzian volume element is given by
\begin{equation}
\sqrt{-g^{F}} = \left(\frac{F}{\alpha}\right)^{5}\left(\frac{S}{\sigma}\right)^{2}\sqrt{-g}.
\end{equation}

\section{Finslerian Einstein-Hilbert action}

Assuming the dynamics of the space-time is governed by an Einstein-Hilbert action
then,
\begin{eqnarray}
\label{bipartitevolumeelement}
 S_{F} & = & \kappa\int{R^{F}\sqrt{-g^{F}}d^{4}xd^{4}y}\nonumber\\
       & = & \kappa\int{R\sqrt{-g}d^{4}x} + \kappa_{F}\int{8b^{2}R\sqrt{-g}d^{4}x}\nonumber\\
       & + & \kappa\int{(4+b^{2}l_{P}^{2})s^{\mu\nu}R_{\mu\nu}d^{4}x} + ....
\end{eqnarray}
it is possible to regain some interaction terms of the bumblebee model action (\ref{bumblebeeactioninteraction}). As a perspective we expect to obtain the dynamical terms of the bumblebee model and some other interaction terms as well.


\section{Acknowledgements}

We are grateful to Alan Kosteleck\'{y} and Yuri Bonder for useful discussions and to the Graduate Physics Program of the Cear\'{a} Federal University and CNPq (National Council for Scientific and Technologic Developing) for financial supporting.


\begin{thebibliography}{99}

\bibitem{Colladay:1996iz}
  D.~Colladay and V.~A.~Kostelecky,
  Phys.\ Rev.\ D {\bf 55}, 6760 (1997).

\bibitem{Kostelecky:2003fs}
  V.~A.~Kostelecky,
  Phys.\ Rev.\ D {\bf 69}, 105009 (2004).

\bibitem{Maluf:2013nva}
  R.~V.~Maluf, V.~Santos, W.~T.~Cruz and C.~A.~S.~Almeida,
  Phys.\ Rev.\ D {\bf 88}, 025005 (2013)



\bibitem{chern}
D. Bao, S. Chern, Z. Shen, An introduction to Riemann-Finsler geometry,
Springer, 1991.

\bibitem{Pfeifer:2011xi}
  C.~Pfeifer and M.~N.~R.~Wohlfarth,
  Phys.\ Rev.\ D {\bf 85}, 064009 (2012)

 \bibitem{Kostelecky:2010hs}
   A.~V.~Kostelecky and N.~Russell,
   Phys.\ Lett.\ B {\bf 693}, 443 (2010).


\bibitem{Kostelecky:2011qz}
  A.~Kostelecky,
  Phys.\ Lett.\ B {\bf 701}, 137 (2011).



\bibitem{Kostelecky:2012ac}
  V.~A.~Kostelecky, N.~Russell and R.~Tso,
  Phys.\ Lett.\ B {\bf 716}, 470 (2012).


\end{thebibliography}
\end{document}